\documentclass{ws-procs961x669}
\usepackage{natbib}
\usepackage{amsmath}
\usepackage{amssymb}
\usepackage{tabularx}
\usepackage{graphicx}
\begin{document}

\wstoc{For proceedings editors: Combining contributions\\ using
WS-procs961x669 master document in \LaTeX2e}{F. Author}

\title{Galactic center G objects as dust-enshrouded stars near the supermassive black hole}

\author{Michal Zaja\v{c}ek and Monika Pikhartová}
%\index{author}{Michal Zaja\v{c}ek and Monika Pikhartová} % or \aindx{Author, F.}

\address{Department of Theoretical Physics and Astrophysics, Faculty of Science, Masaryk University, Kotlá\v{r}ská 2, CZ-611 37 Brno, Czech Republic  \\E-mail: zajacek@physics.muni.cz}

\author{Florian Peissker}
%\index{author}{Michal Zaja\v{c}ek and Monika Pikhartová} % or \aindx{Author, F.}

\address{I.Physikalisches Institut der Universität zu Köln, Zülpicher Str. 77, D-50937 Köln, Germany}

\begin{abstract}
In this contribution, we revisit the model of a dust-enshrouded star orbiting a low-luminosity galactic nucleus (Zajacek et al., 2014, 2016, 2017). Although it is quite challenging for dust to survive in hot X-ray-emitting plasma surrounding supermassive black holes (SMBHs), we now have an observational evidence that compact dusty objects or ``G'' objects can approach the SMBH in the Galactic center (Sgr A*) on the scale of a few 1000 gravitational radii. Since there are about ten G objects in the Galactic center, it is more likely that they are dust-enshrouded stars whose gaseous-dusty envelopes are stable within the corresponding tidal (Hill) radii of the order of a few astronomical units. Such a length-scale is consistent with their infrared broad-band spectral energy distributions. Broad emission lines, in particular Br$\gamma$ recombination line, can be interpreted to arise within the accretion stream from the circumstellar envelopes forming a compact disc that is truncated by the stellar magnetic field. Alternatively, they could also be associated with circumstellar accretion-disc outflows as well as the material within a denser bow shock ahead of the star. In comparison with the line origin in the photoionized envelopes that can generally be tidally stretched, the scenario involving the circumstellar accretion-disc inflow or outflow can ensure that the line luminosity is rather stable, except for the viewing-angle effects. We speculate about the origin of dust-enshrouded stars that could be young stellar objects or binary-merger products.
\end{abstract}


\section{Introduction}

The center of the Milky Way is the closest Galactic nucleus to us, about $\sim 100$- to $\sim 500$-times closer than the closest extragalactic nuclei, such as the low-luminosity nucleus of M31 or the galactic nucleus of the dwarf Seyfert galaxy NGC 4395. For this reason, one can resolve out the gaseous features with the linear scale of only $\sim 460$ AU with eighth-meter class telescopes in the near-infrared domain, considering the distance to the Galactic center of $\sim 8.1$ kpc \citep{2017FoPh...47..553E,2022RvMP...94b0501G}. It is also the only galactic nucleus where one can study the motion of individual stars as well as the properties of the Nuclear Stellar Cluster (NSC) as a whole \citep{2014CQGra..31x4007S}. Since the NSC has a comparable size to globular clusters but is many times more massive, it represents one of the densest stellar environments in the Galaxy. It is therefore a unique test bed for dense stellar cluster dynamics, in particular two-body non-resonant relaxation, scalar and vector resonant relaxation, dynamical friction, stellar collisions, stellar binary dynamics under the influence of distant perturbers, and other processes \citep{2017ARA&A..55...17A}. 

There have been several detailed multiwavelength studies of the stellar and gaseous environment, which helped to shed light on the star-formation history of the NSC. It appears that the NSC is largely composed of old, late-type stars, most of which formed about 10 billion years ago \citep{2020A&A...641A.102S}. These stars appear to be relaxed and form a stellar cusp across all magnitudes, except for the brightest red giants, whose density distribution is core-like. The core-like distribution was likely caused by the preferential depletion of bigger red giants due to their collisions with an accretion disc, jet, other stars or compact remnants, and tidal disruption of large stellar envelopes \citep{2020bhns.work..357Z}.  

Star-formation within the NSC is episodic, with the last episode in the last few hundred million years \citep{2011ApJ...741..108P,2020A&A...641A.102S}. Surprisingly, in the innermost 0.5 pc, there are very young massive stars of the spectral type O and B that must have formed just about 5 million years ago. A fraction of those forms a stellar disc-like structure, which implies the formation within a massive star-forming disc \citep{2003ApJ...590L..33L} that underwent fragmentation into stars due to gravitational instability \citep{}\footnote{For a differentially rotating gaseous disc with the gas temperature $T$ and the surface density $\Sigma$, the criterion for gravitational stability can conveniently be expressed using the Toomre's stability criterion $Q=c_{\rm s}\kappa/(\pi G \Sigma)$, where $c_{\rm s}=\sqrt{\gamma kT/(\mu m_{\rm H})}$ is the sound speed for ideal gas and $\kappa^2=(2\Omega/R)\, \mathrm{d}(R^2\Omega)/\mathrm{d}R$ is the epicyclic frequency of the radially displaced parcel of gas disc whose angular speed at the distance $R$ is $\Omega$. For a Keplerian disc, $\kappa=\Omega$. When $Q<1$ the gaseous disc is gravitationally unstable and fragments into stars, which is typically the case for a warm molecular gas disc when located at $r\gtrsim 0.4\,{\rm pc}$ from Sgr~A*\citep{2004ApJ...604L..45M}.}. The disc could have formed when the gas circularized following the infall of a large molecular cloud \citep{2008ApJ...683L..37W,2008Sci...321.1060B}. The rest of the young stars that are rather scattered could have formed within smaller infalling molecular clouds that were sufficiently dense to undergo star-formation and subsequently, the stellar association underwent tidal disruption and partial capture by Sgr~A*. \citep{2014MNRAS.444.1205J,2016LNP...905..205M,2021ApJ...923...69P}\footnote{Here the criterion for the gravitational collapse of the cloud region with the mass density of $\rho$ and the sound speed of $c_{\rm s}$ is given by the Jeans's length, $\lambda_{\rm J}=c_{\rm s}/\sqrt{G\rho}$. The cloud regions larger than $\lambda_{\rm J}$ undergo gravitational collapse while those that are smaller are stable. However, the classical Jeans's criterion is derived for a homogeneous gravitational potential with no radial gradient, which is not valid in the Galactic center\citep{2013MNRAS.434L..56J}. }.

Despite the proposed modes of star-formation close to the supermassive black hole (SMBH), the occurrence of such a large number of young stars, especially within the inner 0.02 pc from Sgr~A* (S stars of spectral type B), is still considered to be a ``paradox of youth''\citep{2003ApJ...586L.127G}. Strong tidal field in the vicinity of the SMBH as well as intense UV, X-ray, and $\gamma$-ray radiation field should in principle inhibit star-formation \citep{1993ApJ...408..496M}. In addition to the \textit{in situ} formation scenarios, \textit{rejuvenation models} have been proposed, specifically the accretion of fresh hydrogen by stars embedded within the accretion disc \citep{2021ApJ...910...94C}, stellar mergers \citep{2016MNRAS.460.3494S,2019ApJ...878...58S}, or potentially the ablation of surface layers of late-type stars by the jet that was more active in the past \citep{2020ApJ...903..140Z,2024arXiv240917773K}.

To solve the puzzle of star-formation in the Galactic center, it is crucial to answer whether star-formation, albeit limited, continues even nowadays. There are several candidates of young stellar objects that deserve more observations at higher resolution and sensitivity. Among them are comet-shaped objects located within the inner parsec from the SMBH that often exhibit a prominent infrared excess\citep{2015ApJ...801L..26Y,2019A&A...624A..97P,2023ApJ...944..231P}. Moreover, a population of $\sim 10$ infrared-excess, line-emitting ``G'' sources has been identified within the S cluster \citep{2020Natur.577..337C,2020A&A...634A..35P}, which has triggered a lot of interest in terms of their potential to bring deeper understanding of accretion, star-formation, and dynamical processes in the neighbourhood of Sgr~A*.

\section{Dust-enshrouded ``G'' objects - Short review}

The ``G'' object that caught the most attention was the source G2 \citep{2012Natur.481...51G}, which was discovered moving fast towards Sgr~A*. It exhibited a broad Br$\gamma$ emission line with the line width of several $\sim 100$ km/s as well as a prominent near-infrared excess continuum emission that indicated the dust temperature of $\sim 500-600\,{\rm K}$. Because of the detected velocity gradient along G2 and its apparent increase towards the pericenter, which occurred in early 2014, the scenario of a core-less cloud or a star with an extended envelope was preferred. If G2 hosts a star of $\sim 1\,M_{\odot}$, any material located outside its tidal (Hill) radius will be affected by the SMBH gravitational field and will become tidally perturbed. This critical tidal radius at the pericenter of the orbit\citep{2021ApJ...923...69P,2023ApJ...943..183P} is,
\begin{align}
   r_{\rm H}&=a(1-e)\left(\frac{m_{\star}}{3M_{\bullet}} \right)^{1/3}\,\notag\\
   &\simeq 0.6\,\left(\frac{r_{\rm p}}{131.5\,{\rm AU}} \right)\left(\frac{m_{\star}}{1\,M_{\odot}} \right)^{1/3}\left(\frac{M_{\bullet}}{4\times 10^6\,M_{\odot}} \right)^{-1/3}{\rm AU}\,,
   \label{eq_Hill_radius}
\end{align}
where $r_{\rm p}=a(1-e)\approx 131.5\,{\rm AU}$ is the pericenter distance for the latest orbital elements of the G2 object ($a=17.23\,{\rm mpc}$, $e=0.963$)\citep{2023ApJ...943..183P}, $m_{\star}$ is the stellar mass, and $M_{\bullet}$ is the SMBH mass.

In contrast to Gillessen et al.\citep{2012Natur.481...51G}, no significant gradient and changes were found for the Br$\gamma$ emission line by Valencia et al.\citep{2015ApJ...800..125V}. Shortly before the pericenter, the Br$\gamma$ line was redshifted, while after the pericenter, it was blueshifted with comparable line width as well as the same luminosity within the uncertainties. The constant luminosity of the Br$\gamma$ emission line was a rather big surprise since the predictions for a cloud model as well as the bow-shock emission indicated an increase towards the pericenter\citep{2015ApJ...798..111P} and a subsequent decrease\citep{2017ApJ...843...29M}. In addition, Witzel et al. \citep{2014ApJ...796L...8W} detected the G2 source close to the pericenter in 2014 and it was consistent with being a point source in the infrared $L$ band. At the pericenter, the foreshortening factor is close to unity, which implies that we should see the true size of the source. Since the point-spread function for eight-meter class telescopes limits the G2 size to $\lesssim 635\,{\rm AU}$ and no tidal stretching was detected, the intrinsic size of G2 should approach $2r_{\rm H}\sim 1\,{\rm AU}$.

Given the G2 infrared excess and a broad recombination line of hydrogen, several scenarios emerged that have attempted to address these properties as well as the apparent emission stability of the source. The common property of these otherwise quite different models is the presence of a gaseous-dusty envelope. We summarize the main proposed scenarios below in several points, which is an update of Table 1 originally published in Zaja\v{c}ek et al.\citep{2017A&A...602A.121Z}:
\begin{itemize}
    \item {\bf dust-enshrouded star/young stellar object}: Murray-Clay \& Loeb (2012)\citep{2012NatCo...3.1049M}; Eckart et al. (2013)\citep{2013A&A...551A..18E}; Scoville \& Burkert (2013)\citep{2013ApJ...768..108S}; Ballone et al. (2013)\citep{2013ApJ...776...13B}; Zaja\v{c}ek et al. (2014)\citep{2014A&A...565A..17Z}; De Colle et al. (2014)\citep{2014ApJ...789L..33D}; Valencia-S. et al. (2015)\citep{2015ApJ...800..125V}; Ballone et al. (2016)\citep{2016ApJ...819L..28B}; Shahzamanian et al. (2016)\citep{2016A&A...593A.131S}; Zaja\v{c}ek et al. (2017)\citep{2017A&A...602A.121Z}; Morsony et al. (2017)\citep{2017ApJ...843...29M}; Ballone et al. (2018)\citep{2018MNRAS.479.5288B}; Owen \& Lin (2023)\citep{2023MNRAS.519..397O} ,
   \item  {\bf binary/binary merger product}: Zaja\v{c}ek et al. (2014)\citep{2014A&A...565A..17Z}; Prodan et al. (2015)\citep{2015ApJ...799..118P}; Witzel et al. (2014)\citep{2014ApJ...796L...8W}; Stephan et al. (2016)\citep{2016MNRAS.460.3494S}; Witzel et al. (2017)\citep{2017ApJ...847...80W}; Stephan et al.(2019)\citep{2019ApJ...878...58S},
   \item  {\bf core-less cloud/streamer}: Gillessen et al. (2012)\citep{2012Natur.481...51G}; Burkert et al. (2012)\citep{2012ApJ...750...58B}; Schartmann et al. (2012)\citep{2012ApJ...755..155S}; Shcherbakov (2014)\citep{2014ApJ...783...31S}; Pfuhl et al. (2015)\citep{2015ApJ...798..111P}; Schartmann et al. (2015)\citep{2015ApJ...811..155S}; McCourt et al. (2015)\citep{2015MNRAS.449....2M}; McCourt \& Madigan (2016)\citep{2016MNRAS.455.2187M}; Madigan et al. (2017)\citep{2017MNRAS.465.2310M}; Morsony et al. (2017)\citep{2017ApJ...843...29M}; Plewa et al. (2017)\citep{2017ApJ...840...50P}; Gillessen et al. (2019)\citep{2019ApJ...871..126G},
   \item  {\bf tidal disruption event}: Miralda-Escudé (2012)\citep{2012ApJ...756...86M}; Guillochon et al. (2014)\citep{2014ApJ...786L..12G},
   \item  {\bf nova outburst}: Meyer \& Meyer-Hofmeister (2012)\citep{2012A&A...546L...2M},
   \item  {\bf scenarios involving a planet/protoplanet}: Mapelli \& Ripamonti (2015)\citep{2015ApJ...806..197M}; Trani et al. (2016)\citep{2016ApJ...831...61T}; Owen \& Lin (2023)\citep{2023MNRAS.519..397O}.      
\end{itemize}

Some studies, such as the one by Morsony et al.\citep{2017ApJ...843...29M}, are included both in the dust-enshrouded star model as well as in the core-less cloud/streamer scenarios, since they analyze both the stellar compact core as well as the large, extended gaseous-dusty envelope. From the above-mentioned scenarios, the dust-enshrouded star orbiting the SMBH on an eccentric orbit, seems to capture the main observed features of ``G'' objects the best as we discuss below in more detail.

\section{``G'' objects as dust-enshrouded stars close to Sgr~A*}

\begin{figure}
    \centering
    \includegraphics[width=\textwidth]{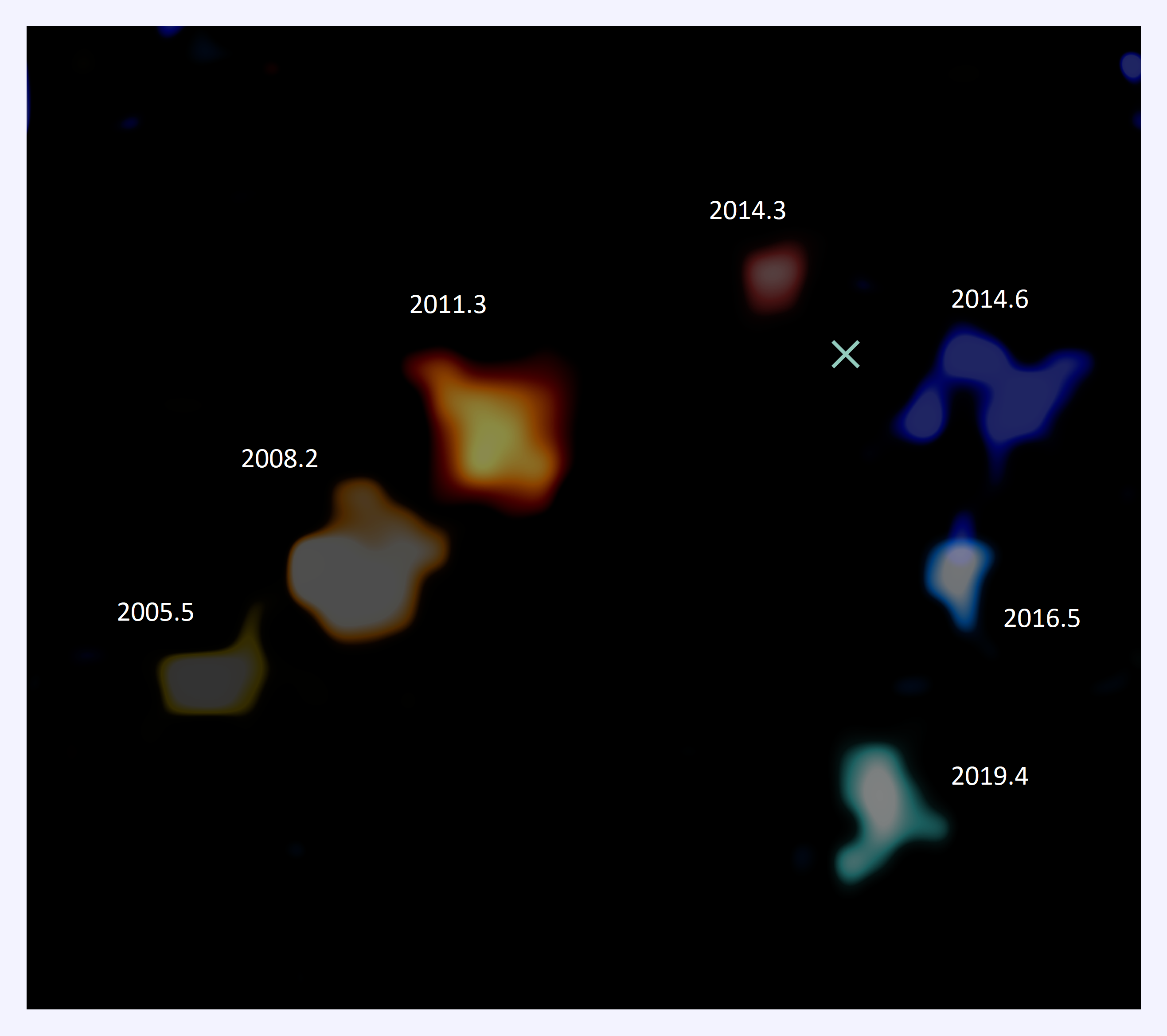}
    \caption{An illustrative temporal evolution of the Br$\gamma$ emission of the G2 object as it moves around Sgr~A* between 2005.5 and 2019.4. The reddened line maps correspond to the pre-pericenter part of the orbit, while the bluish line maps represent the post-pericenter evolution after 2014.4. The blue cross marks the position of Sgr~A*. The north is up, the east is to the left.}
    \label{fig_G2_evolution}
\end{figure}

\begin{figure}
    \centering
    \includegraphics[width=\textwidth]{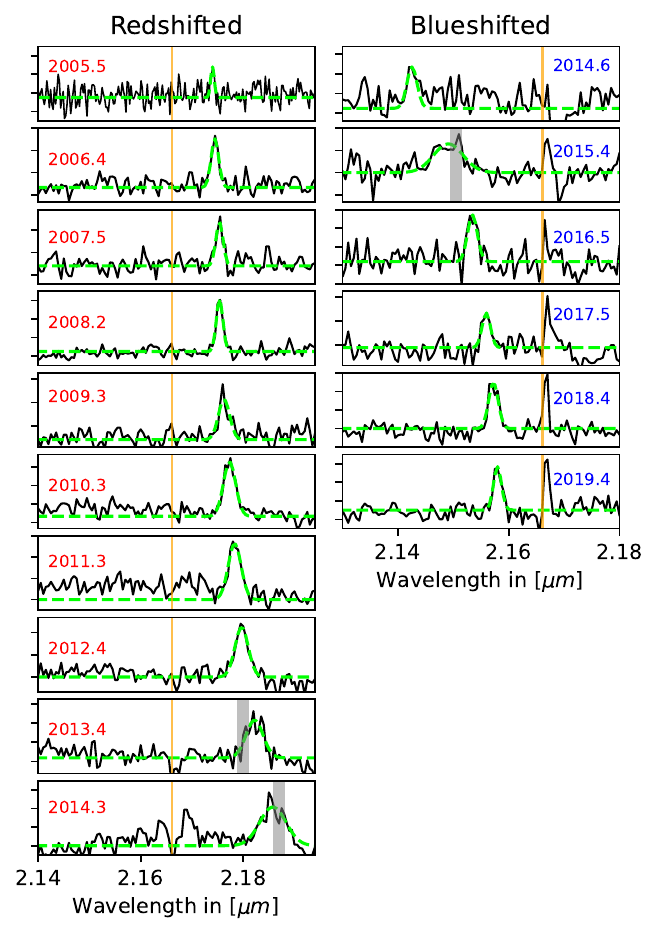}
    \caption{Temporal evolution of the Br$\gamma$ emission line associated with the G2 object orbiting around Sgr~A* on a highly eccentric orbit between 2005.5 and 2019.4. The object reached its periapsis around 2014.4. The gray-shaded rectangle marks the contribution of the telluric OH emission, which can enhance the line width of the Br$\gamma$ line at certain epochs.}
    \label{fig_G2_line_evolution}
\end{figure}

To compare the theoretical predictions of a dust-enshrouded star model with the near-infrared observations of the G2 object and other ``G'' sources, we use the model of a dust-enshrouded star orbiting a low-luminosity galactic nucleus as studied by Zaja\v{c}ek et al. (2014)\citep{2014A&A...565A..17Z}. Further extensions of the model were presented in Zaja\v{c}ek et al. (2016)\citep{2016MNRAS.455.1257Z} and Zaja\v{c}ek et al. (2017)\citep{2017A&A...602A.121Z}. The main characteristics that point towards the dust-enshrouded star model, as outlined for the intensively monitored G2 object, are
\begin{itemize}
    \item the orbit of the G2 object that passed the pericenter around 2014.4 does not deviate significantly from a pure Keplerian orbit around Sgr~A* between 2005.5 and 2019.4\citep{2021ApJ...923...69P}. This is also illustrated in Fig.~\ref{fig_G2_evolution}, which shows the G2 object (its Br$\gamma$ emission) on the sky as it moves around Sgr~A*,
    \item the infrared excess between $K_{\rm s}$ and $L'$ bands (2.2 and 3.7\,$\mu{\rm m}$) can be attributed to a compact dusty envelope with the outer radius of $\sim 5\,{\rm AU}$ and the density corresponding to the accretion rate of $5\times 10^{-7}\,M_{\odot}{\rm yr^{-1}}$ within the infalling, rotationally flattened envelope of an Ulrich type\citep{1976ApJ...210..377U,2017A&A...602A.121Z},
    \item such a compact dusty envelope is not significantly perturbed by the tidal field of Sgr~A*; see also Eq.~\eqref{eq_Hill_radius}. This is consistent with the stable profile of the Br$\gamma$ emission line, which is first redshifted as the source moves towards the pericenter and then blueshifted. There is no significant trace of the simultaneous redshifted and the blueshifted Br$\gamma$ emission as expected for a tidally stretched, ``spaghettified'' material, see also Fig.~\ref{fig_G2_line_evolution},
    \item the relatively large line width of the Br$\gamma$ line with the FWHM of $\sim 100\,{\rm km\,s^{-1}}$ can be explained by the model of magnetospherical accretion, where the inner accretion disc surrounding a young stellar object is truncated at a magnetospheric radius due to the magnetic field pressure of the stellar magnetosphere. The ionized gaseous material at the inner edge of the disc is then channelled along magnetic field lines towards the stellar magnetic poles, where it gets shocked. Along the way, it emits lines such as Br$\gamma$ that get significantly Doppler-broadened due to nearly free-fall motion\citep{2015ApJ...800..125V}. In addition to the inflowing gas, there can also be contributions from disc outflows and the bow-shock emission. Altogether, the line luminosity in this model is intrinsic to the star and is correlated with the accretion rate of gas onto the stellar surface. There can be small variations in the line luminosity that are related to the changing viewing angle as the star surrounded by an axially symmetric structure orbits the SMBH on an inclined elliptical orbit, 
    \item the reported significant polarized emission in the $K_{\rm s}$ band\citep{2016A&A...593A.131S} with the polarization degree of $\sim 20-30\%$ can be achieved when the dusty envelope is significantly non-spherical, such as due to the formation of the bow shock and the presence of bipolar cavities\citep{2017A&A...602A.121Z}. The polarized emission at $2.2\,{\rm \mu m}$ is then attributed to the scattered dust emission,
    \item the near- to mid-infrared spectral energy distribution (SED) of the G2 source, including $H$, $K_{\rm s}$, $L$, and $M$ bands, is fitted better with a two-component SED, consisting of a star and a cooler dusty envelope, than with a one-component SED \citep{2020A&A...634A..35P}.
\end{itemize}

\section{Origin and Formation channels}

Given the points above, we interpret the G2 object as well as other compact structures with similar characteristics as dust-enshrouded stars. Since there is a population of $\sim 10$ such objects within the S cluster\citep{2020Natur.577..337C,2020A&A...634A..35P}, it is challenging to explain their presence using the scenario of a core-less dusty cloud as originally proposed. This is mainly due to very short hydrodynamical timescales of the order of $\sim 10$ years related to the evaporation of the colder material embedded within the hot X-ray emitting plasma as well as to the Kelvin-Helmholtz instability due to the velocity shear as the cloud moves on an eccentric orbit around the SMBH\citep{2012ApJ...750...58B}.  

For a dust-enshrouded star model, two formation mechanisms are plausible:
\begin{itemize}
    \item[(i)] star-formation within the last $\sim 1$ Myr in the close vicinity of Sgr~A* since ``G'' objects share the SED slope with class 1 young stellar objects\citep{},
    \item[(ii)] mergers of binary stars that result in merger products with similar properties as young stars.
\end{itemize}

The likelihood of star-formation process (i) is determined by the availability of cold molecular material close to the sphere of influence of Sgr~A*. Jalali et al.\citep{2014MNRAS.444.1205J} suggested that clump-clump collisions within the circumnuclear disc (CND) could occasionally bring a molecular material towards Sgr~A*. This process is highly uncertain as the clumpiness of the CND is not detemined in detail and the clump-clump collisions may not always result in radially infalling molecular clouds. However, the restrictions on the minimum cloud density to resist destruction by tides are much lower than in the S cluster, which is given by the Roche-limit number density,
\begin{align}
    n_{\rm Roche}&\gtrsim \frac{9M_{\bullet}}{4\pi \mu m_{\rm H}r^3}\,\notag\\
    &\sim 1.2 \times 10^8 \left(\frac{M_{\bullet}}{4\times 10^6\,M_{\odot}} \right) \left(\frac{r}{1\,{\rm pc}} \right)^{-3}\,{\rm cm^{-3}}\,,
\end{align}
which is close to the typical densities of $\sim 0.05-0.2\,{\rm pc}$ clumps within the CND. When they collide, the angular momentum is removed and the cloud starts falling in. Shock waves and intense radiation may provide additional compression to trigger star-formation. A typical timescale to form a star is given by the free-fall timescale, $t_{\rm ff}\sim [3\pi/(32 G \mu m_{\rm H}n_{\rm Roche})]^{1/2}\sim 5150$ years. For the stellar association to form during the clump infall from the CND region at the distance $d_{\rm cl}$, we require that the infall timescale (half of the orbital timescale) is comparable to or longer than the free-fall timescale within the clump. This gives the restriction on the infalling clump initial distance,
\begin{align}
   d_{\rm cl}&\gtrsim  \left(\frac{3M_{\bullet}}{32 \pi \mu m_{\rm H}n_{\rm Roche}} \right)^{1/3}\,\notag\\
   &\sim 0.36\,\left(\frac{M_{\bullet}}{4\times 10^6\,M_{\odot}} \right)^{1/3}\left(\frac{n_{\rm Roche}}{10^8\,{\rm cm^{-3}} }\right)^{-1/3}{\rm pc}\,,
\end{align}
which is indeed at the inner edge of the CND and still within the sphere of influence of the SMBH.   
Despite the difficulties, if such a cloud falls in towards the SMBH at least once per million years and it reaches the critical density for the gravitational instability at the site of collision, then the timescale for the cloud to reach Sgr~A* is comparable to the free-fall timescale of protostellar fragments to form stars\citep{2021ApJ...923...69P}. 

The binary merger (ii) likelihood depends on the typical timescale, during which the components can merge in the vicinity of Sgr~A*. This is approximately given by the Kozai-Lidov timescale for the period-inclination oscillations driven by the SMBH\citep{2016MNRAS.460.3494S,2019ApJ...878...58S},
\begin{align}
    \tau_{\rm KL}^{\rm SMBH}&=2\pi \left(\frac{m_{\rm bin}}{M_{\bullet}} \right) \left(\frac{a_{\rm G}}{a_{\rm bin}} \right)^3 P_{\rm bin}\,\notag\\
    &\simeq 8.6\times 10^5 \left(\frac{m_{\rm bin}}{2\,M_{\odot}} \right) \left(\frac{M_{\bullet}}{4\times 10^6\,M_{\odot}} \right)^{-1} \left( \frac{a_{\rm G}}{0.04\,{\rm pc}}\right)^3 \times\,\notag\\
    &\times \left( \frac{a_{\rm bin}}{1.25\,{\rm AU}}\right)^{-3} \left(\frac{P_{\rm bin}}{1.0\,{\rm years}} \right)\,\text{yr}\,,
\end{align}
where $m_{\rm bin}$, $a_{\rm bin}$, and $P_{\rm bin}$ are the binary mass, semi-major axis of the components, and the binary period, respectively, while $a_{\rm G}$ is the semi-major axis of the binary (``G'' object precursor) around the SMBH. The Kozai-Lidov timescale $\tau_{\rm KL}$ is thus shorter than the estimated lifetime of S stars and OB stars, which again implies the young dynamical age of ``G'' objects. 
Another parameter that is relevant for this channel is the binarity fraction, which is uncertain in the Galactic center environment, especially close to Sgr~A* within the S cluster. Specifically, there appears to be radial dependence of the binary star fraction close to the Galactic center \citep{2024ApJ...964..164G}, with the trend of a decreasing binary fraction towards Sgr~A* \citep{2023ApJ...948...94C} due to dynamical processes of binary-component mergers or disruptions in the vicinity of the SMBH. This should, however, be confirmed for fainter stars when 30/40-meter class telescopes with a larger sensitivity are available. 

\section{Conclusions}

Depending on whether the process (i) or (ii) operates on a shorter timescale in combination with the final number of produced stars with G2-like properties, it is then the dominant contributor to the observed population of ``G'' objects. Future NIR- and MIR-observations of the S cluster region with more sensititve 30-/40-meter class telescopes will likely reveal more ``G'' objects towards lower luminosities/masses, which will be instrumental in constraining the origin of these mysterious stellar objects. Another important parameter is the binarity fraction among the S stars. Detection of binaries can significantly constrain dynamical mechanisms, in particular their timescales and the involved perturbing mass, operating deep in the potential well of Sgr~A*. In spite of many open questions, G objects have shown the way how cold material, in particular dust, can be transported to the distance of only $\sim 100\,{\rm AU}$, which can be applied to improve the models of hot accretion flows at intermediate scales\citep{refId0}.

\section*{Acknowledgements}

We thank the editor for useful comments that improved this manuscript. MZ and MP acknowledge the support from the Czech Science Foundation Junior Star grant no. GM24-10599M.

%\begin{thebibliography}{9}
%\bibitem{lamp94}
%L.~Lamport, \emph{\LaTeX, A Document Preparation System}. (Addison-Wesley,
%  Reading, MA, 1994), 2nd edition.
%\bibitem{lamp87} L. Lamport, \emph{Make Index: An Index Processor For LaTeX}, (1987).
%\end{thebibliography}

%\bibliographystyle{abbrv}
%\bibliography{references}

\end{document}